\documentclass[preprint]{elsarticle}

\usepackage{graphicx}

\usepackage[cmex10]{amsmath}
\usepackage{array}
\usepackage[utf8]{inputenc}

\begin{document}

\begin{frontmatter}

\title{Tasks for Agent-Based Negotiation Teams: Analysis, Review, and Challenges}

\author[upv]{Victor Sanchez-Anguix} \ead{sanguix@dsic.upv.es}
\author[upv]{Vicente Julian} \ead{vinglada@dsic.upv.es}
\author[upv]{Vicente Botti} \ead{vbotti@dsic.upv.es}
\author[upv]{Ana García-Fornes} \ead{agarcia@dsic.upv.es}

\address[upv]{Departamento de Sistemas Informáticos y Computación\\ Universitat Politècnica de València\\ Camí de Vera s/n, 46022, Valencia, Spain}

\begin{abstract}
An agent-based negotiation team is a group of interdependent agents that join together as a single negotiation party due to their shared interests in the negotiation at hand. The reasons to employ an agent-based negotiation team may vary: (i) more computation and parallelization capabilities; (ii) unite agents with different expertise and skills whose joint work makes it possible to tackle complex negotiation domains; (iii) the necessity to represent different stakeholders or different preferences in the same party (e.g., organizations, countries, married couple, etc.). The topic of agent-based negotiation teams has been recently introduced in multi-agent research. Therefore, it is necessary to identify good practices, challenges, and related research that may help in advancing the state-of-the-art in agent-based negotiation teams.   For that reason, in this article we review the tasks to be carried out by agent-based negotiation teams. Each task is analyzed and related with current advances in different research areas. The analysis aims to identify special challenges that may arise due to the particularities of agent-based negotiation teams.
\end{abstract}


\end{frontmatter}

\section{Introduction}
\label{intro}

According to Pruitt \cite{pruitt81}, negotiation can be defined as a process in which a joint decision is made by two or more parties. The parties first verbalize contradictory demands and then move towards agreement by a process of concession-making or search for new alternatives. Therefore, automated negotiation consists of an automated search process for an agreement between two or more parties where participants exchange proposals. 

The work in automated negotiation has taken two alternative and complementary paths: game-theoretic approaches and heuristic approaches. Game theory researchers focus on reaching solutions at equilibrium under varied assumptions like unbounded computational resources, complete information, or some statistical information regarding the strategies and preferences of the other parties \cite{fatima04a,serrano03,gatti11}. 

Unfortunately, some of these assumptions may not be present in some real world applications. For instance, computational resources are of extreme importance for agents since they may be scarce and shared among different tasks (e.g., time, memory usage, etc.). Thus, negotiation should not always assume unbounded computational resources. Additionally, since agents are heterogeneous, not all of the agents know the same strategies. Identifying which sets of strategies are known by each agent may be a hard task that can only be successful after several negotiations. The same goes for the knowledge regarding the opponents' preferences, reservations values, and so forth. Only after several interactions and negotiations with the opponents, the agent may be able to come with an approximate model of the other agents. Hence, models that tackle uncertainty and limit the use of computational resources are mandatory for some situations. Heuristic models usually avoid the aforementioned assumptions. In result, they usually do not find equilibriums, but they obtain satisfactory results \cite{kraus97,jennings01}. From this point, this work assumes the heuristic perspective when computational models for automated negotiation are mentioned.

Most of the research in automated negotiation has concentrated on bilateral bargaining and multi-party negotiations where parties are composed of single individuals. However, most real life negotiations, especially in business and politics, involve negotiation teams \cite{thompson96,peterson97,thompson01}: multi-individual parties that act together as a single negotiation party. The use of negotiation teams in automated negotiation and multi-agent systems is relatively new \cite{sanchez-anguix10,sanchez-anguix13,sanchez-anguix12,sanchez-anguix12b,sanchez-anguix12c}. Hence, a proper review needs to be done in order to identify the tasks and special challenges that may arise in negotiation teams' settings. The main goal of this article is analyzing the tasks to be carried out by agent-based negotiation teams, reviewing how current works may contribute to the consecution of such tasks, and identifying special challenges that may arise due to the particularities of agent-based negotiation teams. We also hope that this present work may help new researchers in the area to have a global perspective on the problems that may arise in agent-based negotiation teams.

The remainder of this paper is organized as follows. In Section \ref{motivation} we start by introducing negotiation teams from the point of view of the social sciences. We identify the main aspects that characterize human negotiation teams and, after that, we relate those aspects with the topic of agent-based negotiation teams. Then, we present an analysis on the tasks to be carried out by agent-based negotiation teams in Section \ref{workflow}. Finally, we conclude by overviewing the work presented in this article.

\section{Motivation}
\label{motivation}
As a first introspective to negotiation teams, we describe the phenomenon from the point of view of the social sciences. Then, we relate the essential aspects identified in the social sciences with agent-based negotiation teams.
\subsection{Human Negotiation Teams}
 A negotiation team is a group of two or more interdependent persons who join together as a single negotiating party because their similar interests and objectives relate to the negotiation, and who are all present at the bargaining table \cite{thompson96,thompson01}. Hence, a negotiation team is a negotiation party that is formed by multiple individuals instead of just one individual. As a negotiation party, the team negotiates with other parties in order to reach  a final agreement. 

In what kind of scenarios may a negotiation be involved? There are several scenarios and their importance range from day to day negotiations to crucial negotiation in real life like business and politics. For instance, we can think of the following negotiation cases where teams usually participate in real life:
\begin{itemize}
 \item In one scenario, a human organization like a company desires to sell a product line to another company \cite{peterson97}. It is usual for each company to send a negotiation team composed of different experts coming from different organizational departments. This team is entrusted with the task of understanding the complex scenario at hand and obtaining the most in the final agreement. 
 \item Similarly to the scenario mentioned above, negotiations in politics also involve negotiation teams. We could for instance think of the negotiations carried out between Cambodia, Laos, Thailand and Vietnam for promoting cooperation on water resources \cite{browder00}. In these negotiations, each national party formed a negotiation team. Each team was formed of different specialists whose common goal was the quality of the final agreement for its country.
 \item Imagine that a married couple wants to purchase a car \cite{peterson97}. For that matter, the couple has to negotiate with a car seller the purchasing conditions like price, payment method, and extras. Clearly, this is an agreement that is signed between two parties: the couple, and the car seller. However, one of the parties is clearly composed of two individuals (i.e., the couple is composed of the husband and the wife) that share the same goal (i.e., buy a car).
 \item Suppose that a group of four friends decides to travel together. If a travel needs to be arranged, the group of friends needs to find an adequate destination, some nice accommodation, and flights that take the group to the city of their holidays. Additionally, it may even be interesting to include some pre-arranged social activities like visits to museums, visits to monuments, some sport activities, and so forth. There may be several travel agencies that offer such services, and the group of friends may need to negotiate with some of them to get a travel package that satisfies their needs. As in the previous scenarios, the group of friends is one single negotiation party that is composed of multiple individuals that share a common objective (i.e., go on a trip together).
\end{itemize}

Thus, it can be appreciated that negotiation teams are quite common in real life negotiations. Despite their importance in real negotiations, teams have not been studied by social sciences to the same extent as dyadic negotiations \cite{connor97,gelfand05,behfar08}. However, what are the reasons to send a negotiation team to the negotiation table instead of a single negotiator? The first reason that may come to our minds is that the more individuals the better task performance in negotiation. Thompson et al. \cite{thompson96} showed in several experiments involving human negotiation teams that as long as one of the parties is a negotiation team, better joint outcomes (i.e., integrative outcomes) were obtained. This is partially explained due to the fact that when teams are present at the negotiation table, parties are more inclined to exchange information \cite{thompson96}. However, as pointed out by Mannix \cite{mannix05}, greater numbers may not pay off unless coordination is present among team members. 

Another reason to employ a negotiation team is skill distribution and information distribution \cite{peterson97,mannix05,thompson05}. With this, we mean that different team members may have different and complementary profiles needed to tackle properly the negotiation problem. Thompson  \cite{thompson05} recommends that managers should recruit negotiation teams composed of experts in negotiation, experts in the subject to be negotiated, and individuals with a variety of interpersonal skills. Mannix \cite{mannix05} states that negotiation teams require a diverse set of knowledge, abilities, or expertise in complex negotiations, and the correct assessment of such skills is pointed out as one of the keys for success in a negotiation. Skill distribution and complementary skills are of vital importance when using some classic team negotiation tactics like the good cop/bad cop persuasion tactic \cite{brodt00}.

Finally, other authors consider that another reason to send a negotiation team to the negotiation table is the party being formed by different stakeholders \cite{mannix05,halevy08}. For instance, Mannix \cite{mannix05} points out union negotiation as an example of negotiations where parties are formed by different interests that have to be represented in the negotiation table. Halevy \cite{halevy08} also remarks the importance that despite negotiation teams being a single negotiation party, they are hardly ever a unitary player. In fact, a negotiation team is usually a multi-player party with different and possibly conflicting interests. 

\subsection{Agent-Based Negotiation Teams}

Up to this point, we have strictly considered human negotiation teams. However, are negotiation teams feasible and needed for automated negotiation and electronic applications? We argue that the answer to such question is positive. Agent-based systems are not alien to negotiation scenarios where it may be interesting to employ negotiation teams. For instance, imagine a tourism e-market application. It is usual for groups of friends/families, or even strangers, to organize their holidays as a group. Nevertheless, travelers usually have different preferences regarding trip conditions (e.g., cities to visit, hotel location, leisure activities, number of days to spend, budget limitations, etc.). Humans may be slow at coming with a proper negotiated  deal that accounts for everyone's preferences. This task may not be only slow, but also tedious since conflict may be present.  On top of that, human negotiators have been documented to be content even with deals that are far from the optimal solution. This phenomenon is what is known in the literature as \textit{leaving money on the table} \cite{thompson05}.  As a way for overcoming these problems, agents representing each traveler could form a team in an electronic market to quickly obtain a good trip package for the group. Software agents should be quicker than humans as long as they can handle the negotiation domain and they should prevent human negotiators from carrying out tedious tasks.  The application of negotiation teams is not limited to the aforementioned example. It can be extrapolated to other domains such as electronic farming cooperatives, customer coalitions, organizational merging between virtual organizations, negotiation support systems for labor negotiations, and so forth. 

A trusted mediator who can perfectly aggregate preferences can be thought of as a possible mechanism to coordinate a negotiation team in the tourism e-market example. Nevertheless, there are several reasons that preclude us from aligning ourselves with this kind of coordination mechanisms. The first reason is that privacy is usually a concern among users in electronic applications. In fact approximately 90\% of the users in electronic applications care to some degree about the amount of information that they filtrate in electronic applications, and only 10\% do not care about letting others manage their information \cite{taylor03}. Hence, one cannot expect that every team member may be willing to share its full preferences to a mediator. The other important issue is the fact that even though there may be some degree of cooperation among team members, one should not forget that the team is a multi-player party and opportunistic behavior may be present. In that case, preference aggregation is a dangerous mechanism since it may be quite prone to being manipulated and exaggerated for one's own benefits. Therefore, new negotiation models are needed to coordinate agent-based negotiation teams. 

Making an analogy with the human counterpart, we can define an agent-based negotiation team as a group of two or more interdependent agents that join as a single negotiation party because of their shared interests in the negotiation. The reasons to use an agent-based negotiation team are also analogous to the human case. First, more agents in the team may mean more computation capabilities and, thus, more extensive and parallelized exploration of the negotiation space. Second, we can also assume that different and heterogeneous agents may have different experiences, they may offer different services/skills, and they may implement different algorithms, which in the end may result in the team being able to tackle complex negotiation problems more efficiently. Third and lastly, the team may really represent a multi-player party whose preferences need to be satisfied as much as possible by the final agreement. 

Another related topic is one-to-many and many-to-many negotiations \cite{smith80,sandholm93,nguyen04,an06,shoham09,williams12,mansour12}. One-to-many and many-to-many negotiations are concepts that address the number of parties involved in the negotiation. In one-to-many negotiations, one party negotiates simultaneously, in with multiple parties. It can be a party negotiating in parallel negotiation threads for the same good with different opponent parties \cite{nguyen04,an06,williams12,mansour12} or a party that negotiates simultaneously with multiple parties like in the Contract-Net protocol, and the English and Dutch auction \cite{smith80,sandholm93,shoham09}. Many-to-many negotiations consider the fact that many parties negotiate with many parties, the double auction being the most representative example \cite{shoham09}. Differently to the aforementioned concepts, negotiation teams are not related with the cardinality of the parties but the nature of the party itself. When addressing a negotiation team, we consider a negotiation party that is formed by more than a multiple individuals whose preferences have to be represented in the final agreement. This complex negotiation party can participate in bilateral negotiations, one-to-many negotiations, or many-to-many negotiations. The reason to model this complex negotiation party instead of as multiple individual parties is the potential for cooperation. Despite having possibly different individual preferences, a negotiation team usually exists because there is a shared common goal among team members which is of particular importance (e.g., going on a trip together in the case of the group of travelers, selling the product line in the case of the human organization, purchasing a car in the case of the married couple, and obtaining a beneficial agreement for their country/region in the case of politics). 

Nevertheless, human negotiation teams do not always guarantee a better outcome than individuals. The performance of the team is directly related to coordination among team members. A team that is not capable of achieving such coordination may fail at the negotiation. In fact, Behfar et al. \cite{behfar08} study the causes that pose problems for human negotiation teams: logistics and communication problems (e.g., communications inefficiencies), substantive differences (e.g., confusion about goals, conflicting interests), inter-personal and personality differences (e.g., different negotiation styles), and confusion about team roles (e.g., unclear decision rights). The same authors also identify those strategies that help to overcome the aforementioned problems and lead teams towards success: time and logistics management (e.g., coordinating strategies during negotiation by stepping away from the table), team communications (e.g., preparing with teammates), within-team negotiations (e.g., team problem solving, managing conflicting interests), and defining leadership and team roles (e.g., defining decision rights). To put it briefly, communications, coordination, intra-team negotiation, and clear rules of the game lead human negotiation teams to success. We believe that those key elements are also important in agent-based negotiation teams. For that reason, we identify and describe the communications, coordination and negotiation tasks that may help agent-based negotiation teams to perform successfully. Solutions for each task are reviewed, and special challenges that may arise are also pointed out.

This present work can be categorized as a review paper from the point of view of those areas that may be relevant to agent-based negotiation teams. In the automated negotiation literature, one can find different descriptive studies and reviews that have appeared in the last few decades. Kraus \cite{kraus97} presents a review of negotiation and cooperation in multi-agent systems by analyzing works in game-theory, large scale mechanics, operation research, and heuristics. Beam et al. \cite{beam97} presented another point of view of the research in automated negotiation. Their study focused on challenges for electronic commerce and, therefore, the article contained state-of-the-art sections concerning negotiation support systems, automated intelligent agents in negotiation, auction mechanisms, and electronic marketplaces. Later, Guttman et al. \cite{guttman98} overview the use of negotiating agents and agent technologies in electronic marketplaces while putting a special emphasis on some of the most important tasks needed in electronic marketplaces: product brokering, merchant brokering, and negotiation. Similarly to Kraus \cite{kraus97}, Jennings et al. \cite{jennings01} offer a review of some of the most important works in negotiation, advances, and challenges in game-theoretic models, heuristic models, and argumentation frameworks. Another review regarding agents in electronic commerce is presented by Lomuscio et al. \cite{lomuscio03}. The authors present a taxonomy to classify negotiation in electronic commerce attending to criteria like the cardinality of the negotiation (i.e., number of parties involved), the characteristics of the agents (e.g., role, rationality, knowledge, bidding strategy, etc.), the characteristics of the environment, the type of information exchanged, and event parameters. The taxonomy is applied to several well-known negotiation frameworks in the literature. He et al. \cite{he03} review agent-based approaches to electronic commerce from the point of view of business-to-consumer (B2C) and business-to-business (B2B) markets. In the former case, the authors review different approaches in one of the following tasks: need identification, product brokering, buyer coalition formation, merchant brokering, and negotiation. In the latter case, different approaches are reviewed in tasks such as partnership formation, brokering, and negotiation. In \cite{lopes08}, Lopes et al. propose the necessary tasks for the complete implantation of a general negotiation framework. Later, Luo et al. \cite{luo12} propose a software methodology for the design of negotiation models in different scenarios. In the article, they review and standardize different high impact negotiation models. All of the aforementioned works have reviewed different advances and particularities of the research in automated negotiation. However, none of them has focused on the specific problem of negotiation teams and the special challenges associated to the topic.   

\section{Tasks for Negotiation Teams}
\label{workflow}

In this section, we describe the different tasks that an agent-based negotiation team may need to carry out in different scenarios. For each task, a review of related literature is carried out and the special challenges that are present in the negotiation team setting are highlighted.

Some authors \cite{lopes08} have suggested the following tasks for negotiating agents: Identify social conflict, identify negotiation parties, structuring personal information, analysis of the opponents, define a protocol and select a negotiation strategy, negotiation (i.e., exchange of offers, argumentation, learning, etc.), and re-negotiation. We argue that the tasks proposed by other authors are the correct steps to be followed by negotiating parties. However, some additional tasks, mainly those involving team coordination, are introduced to cope with the team scenario. It should be noted that depending on the application domain, some of the tasks described may not be necessary.

\begin{figure}[t]
 \centering
 \includegraphics[width=0.65\linewidth]{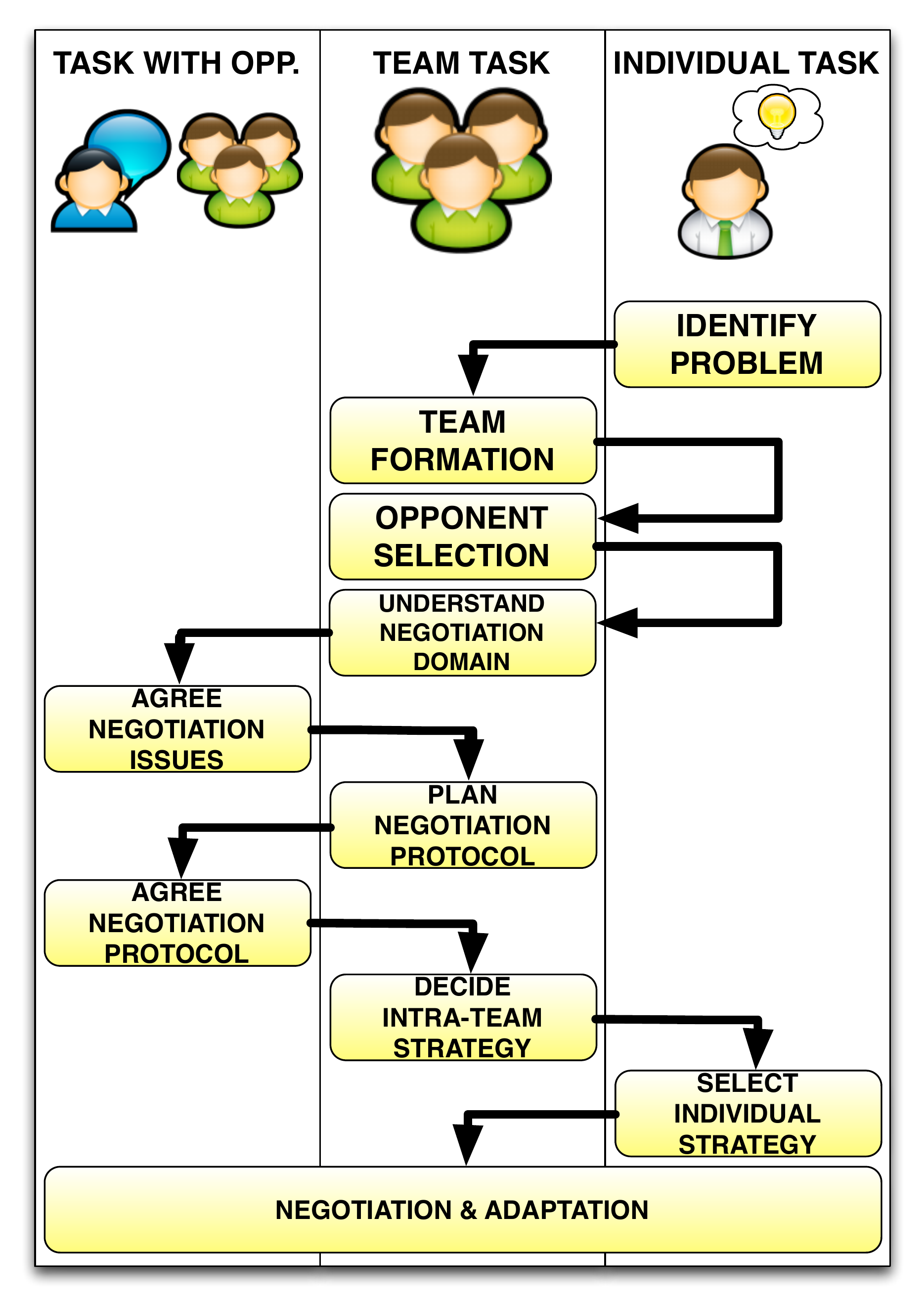}
 \caption{A general workflow of tasks for agent-based negotiation teams.}
\label{scheme}
\end{figure}

The tasks have been organized in a workflow as it can be observed in Figure \ref{scheme}. In the schema, we distinguish between tasks that are carried out with opponents (task with opp.), tasks that mainly concern interactions with team members (team task), tasks that only involve one individual agent (individual task), and tasks that involve team members, opponents, and the individual. The workflow is therefore divided into \textit{Identify Problem} (individual task), \textit{Team Formation} (team task), \textit{Opponent Selection} (team task),  \textit{Understand Negotiation Domain} (team task), \textit{Agree Negotiation Issues} (Task with Opponents), \textit{Plan Negotiation Protocol} (team task), \textit{Agree External Negotiation Protocol} (task with opponents), \textit{Decide Intra-team Strategy} (team task), \textit{Select Individual Strategy} (individual task), and \textit{Negotiation \& Adaptation} (team, individual and opponents). The workflow aims to be as general as possible so that it could be adapted to a wide variety of domains. Hence, some of the tasks proposed may not be necessary and they can be skipped in some of the domains (e.g., selecting an opponent when there is one single opponent and it is known from the domain). Next, we describe all of the tasks in detail.

\subsection{Identify Problem}

The first step consists of identifying a situation that requires negotiation. The agent has to look in its environment for partners, and, in the process, determine whether a conflict exists or not. As stated by Lopes et al. \cite{lopes08}, most artificial intelligence researchers have focused on how to reach an agreement, but very few have studied the problem of detecting conflict. The final product of this task is a list of potential team members, a list of potential opponents, and, possibly, a list of potential competitors. Next, we review different solutions in the literature to this problem.

\subsubsection{Domain Specific Conflict Detection}

In multi-agent literature, one can identify works where conflict detection mechanisms are designed for specific domains like cooperative planning or air traffic management \cite{tomlin98,chu00,barber01,wollkind04}. In \cite{tomlin98,wollkind04}, virtual cylinders delimit the protected and alert zone of each aircraft. Conflict is detected whenever the cylinders of two different aircrafts overlap. Chu et al. \cite{chu00} propose a conflict detection mechanism for collaborative planning between pairs of agents. The agents work towards a shared plan by exchanging plan proposals and beliefs. One agent is able to detect whether or not a conflict exists by checking proposals and beliefs with respect to its own belief base. Similarly, Barber et al. \cite{barber01} propose a conflict detection mechanism when agents integrate their individual plans into a global plan. More specifically, conflict is detected at the goal level (i.e., incompatible goals) or planning level (i.e., threats between actions) by means of the use of PERT diagrams. In other applications like electronic commerce, conflict/negotiation comes inherently linked to the role of the agent: buyers want to buy at lower prices, while sellers want to maximize their profit. However, these mechanisms are specific and cannot be adapted to other potential domains.

\subsubsection{Domain Independent Conflict Detection}
Research into domain independent mechanisms that allow conflict detection is a topic that needs further research. This research is especially important if one attempts to design general negotiators for different domains. Some researchers like Lopes et al. \cite{lopes02} have employed libraries of axioms (i.e., libraries of rules) that allow agents to compare their own plans and intentions with those expected plans and intentions of other agents. A conflict is detected when one's own intentions are not compatible with predicted intentions of other agents. Similarly, it may be possible to express axioms for detecting potential partners. The main disadvantage of this approach is that libraries of axioms are static, usually provided by experts, and they cannot learn from new situations.

\subsubsection{Looking for Partners and Opponents}

Unless the multi-agent system is a closed one, it is necessary to look for potential partners and opponents. The search process may be a necessary step for the aforementioned conflict detection mechanisms. One interesting technology is searching/communicating in social networks and markets \cite{sarne10,delval12}. Sarne et al. \cite{sarne10} present an efficient algorithm for cooperative search in electronic markets. An agent or a group of agents look for goods in the marketplace to fulfill different goals. The algorithm takes into account purchase opportunities (i.e., the probability of finding a certain good) and the cost associated with finding new opportunities. For that matter, the agents employ distributions over opportunities in the electronic market. The search is capable of being partitioned in autonomous searches that aim to cover the different goals of the agent/group. While the search algorithm is interesting due to the fact that it considers search costs, obtaining distributions over partners/opponents for a specific task/negotiation may be difficult, especially if the negotiation task is new for the agent. Del Val et al. \cite{delval12} present a decentralized service discovery system  in social networks. The algorithm for service discovery is not based on previous information or statistics that require training, but similarity between agents. The similarity measure is calculated considering the semantic descriptions of the agents. Similarity measures may be employed to look for negotiation partners and opponents. However, it requires that agents make public certain semantic information regarding their profiles.

\subsubsection*{\textbf{Special Challenges}} Domain independent conflict resolution is a topic that needs further research. Mechanisms proposed in the literature have the disadvantage of being generated according to the knowledge of experts. Thus, they may not be robust to unexpected scenarios unless learning mechanisms are used. One way to address this issue may be case-based reasoning \cite{lopez01}. Instead of using axioms, the agent may have a database of past conflict situations. If the current situation is similar enough to a past conflict situation, it can be considered that conflict may exist in the present. Once unexpected scenarios arise, it is possible for the agent to store the current situation as a new case. In future events, the new information may be used to predict new similar scenarios. The same mechanism may be used for detecting potential partners.

\subsection{Team Formation}

Once the agent has studied which agents may be considered as potential partners, and which agents may be considered as opponents, the agent faces the challenge of determining whether or not benefits arise from forming a negotiation team. In some situations, it may be mandatory for the agent to be part of a negotiation team. In fact, the team may even be static (i.e., a married couple negotiating with a seller over an apartment). If that is the case,  identifying negotiation partners, and forming a negotiation team  are tasks that can be skipped. Nevertheless, some scenarios may be less rigid and the agent may be able to form a negotiation team from the list of potential partners. Thus, the agent should analyze which team he expects to be the optimal negotiation team according to the list of potential partners. If it is expected that no team reports more benefits for the agent than negotiating individually, the agent should decide to negotiate as a single individual party. Next, we review those mechanisms that have been classically used to distribute agents into groups.

\subsubsection{Coalition Formation}
Allocating agents into optimal groups has been a field of study for coalition formation \cite{sandhlom97,conitzer04,ohta08,zick11,rahwan11}. A coalition is a short-lived group of agents that joins together for the accomplishment of a certain goal. The problem of coalition formation is usually divided into three main problems \cite{rahwan07}: coalitional value calculation, coalition structure generation, and payoff distribution. The first problem, coalitional value calculation, consists of computing every possible coalition of agents and its associated value. Once every possible coalitional value is computed, the next task consists of dividing agents into disjoint coalitions that maximize the social welfare of the system. Finally, the payoff generated by the coalition is divided among the agents that are part of the group. 

In a team negotiation process, payoffs may be difficult to anticipate since the outcome of a negotiation is uncertain. On top of that, the result of the negotiation may be an object whose payoff may not be divided among team members. For instance, in the case of the traveling friends, the final result of the negotiation is the travel. Even though the cost of the travel may be divided among team members, how do you expect to divide the benefits of other factors of the negotiation such as the payment method, arranged foods, hotel location, and so forth? Thus, those works that focus on optimally dividing payoffs \cite{conitzer04,ohta08,zick11} may not be so interesting from the point of view of agent-based negotiation teams. 

On the other hand, efficiently calculating coalitional values and coalition structures \cite{sandhlom97,rahwan09,rahwan11} may come handy in negotiation team formation. This is especially true if the formation task is supported by some kind of centralized authority (i.e., a mediator, the electronic commerce site, etc.). Sandhlom et al. \cite{sandhlom97} present a coalition formation algorithm for problems where computing the cost of the coalitional value is exponential and agents have to pay for resources associated to the computation. Thus, as less computational resources are dedicated to the calculation of the coalitional value, less optimal solutions are obtained. The setting is similar in negotiation team formation. The exact value of the coalitional value cannot be calculated since it is uncertain and the actual value depends on the result of the negotiation. Therefore, approximated methods or heuristics should be used to approximately assess the quality of a coalition. In \cite{sanchez-anguix13}, results suggest that team performance is usually improved as team members' preferences are more similar. Thus, methods like similarity heuristics may be useful to calculate coalitional values when negotiation teams are concerned. Rahwan et al. \cite{rahwan09} present a distributed anytime algorithm for computing coalitional structures. The main advantage of this approach comes from the fact that despite being an anytime algorithm, it can guarantee a bounded solution with respect to optimality. This feature makes it applicable for negotiation teams in marketplaces, where the number of potential partners may be too large for non-anytime methods. Another relevant work for agent-based negotiation teams is constrained coalition formation \cite{rahwan11}. In this framework, not every coalition is possible due to constraints imposed by the application domain. Therefore, one should avoid generating and using coalitions which are not allowed. For instance, in a group travel electronic market, some agents may not be willing to travel with other agents (e.g., due to past experiences).

Coalition formation algorithms are usually (not necessarily always) employed in a semi-centralized way, even when calculations can be distributed. These algorithms are usually applied by a centralized authority that manages all of the coalitional needs of the agents in the system. The authority distributes all the agents in coalitions with the aim of maximizing the social welfare of the system, which may be different to the social welfare of individual agents. Thus, this kind of mechanism is usually employed in settings where such a central authority exists (e.g., an electronic marketplace).

\subsubsection{Buyer Coalitions}

Another trend of research in coalition formation are buyer coalitions \cite{tsvetovat01,yamamoto01,ito02,li03,he06}: groups of buyers that join together in order to take advantage from volume discounts. In \cite{tsvetovat01}, the authors show the incentive for buyers to form coalitions that take advantage of volume discounts. Additionally, they discuss the advantages and disadvantages of different protocols for the formation of customer coalitions (i.e., pre-negotiation and post-negotiation). 

Yamamoto et al. \cite{yamamoto01} present an algorithm for customer coalition formation in marketplaces where agents are able to make bids on different items. A mediator gathers the bids (i.e., price) and groups together buyers that prefer the same item. Then, the surplus is divided fairly inside the coalition so that the coalition is stable (i.e., agents do not have incentive to leave the coalition and join another one). Later, Li et al. \cite{li03} study the problem of forming customer coalitions and distributing the surplus. They study the problem from the point of view of incentive compatibility (i.e., whether buyers have incentive to express their real reservation prices) and coalition stability. Several mechanisms are proposed for the problem, and experiments show a positive correlation between incentive compatibility and coalition stability.

Ito et al. \cite{ito02} present customer coalition models for agents that participate in a marketplace where buyers with the same needs are grouped together and they express their purchase desires in terms of reservation prices. On the other hand, sellers have deadlines to sell their products and they can cooperate with other sellers in case that stock requirements are not met. When a seller's deadline is reached, the seller negotiates with the coalition and buyers can decide whether or not to purchase the product. 

In \cite{he06}, the authors propose a customer coalition formation mechanism for marketplaces where agents can buy bundles of items. An empty group is virtually formed for each seller. Each buyer agent searches for the optimal bundle that it can get by itself. The agent can register in those groups that are part of its optimal bundle. Then, buyer agents negotiate among them to form coalitions that take the most from the purchase options.

In general, the relationship between buyer coalitions and negotiation teams is strong: both are groups that engage in some kind of negotiation as a single party. The main disadvantage of current buyer coalition approaches is the fact that most works in group buying have focused on single attribute transactions where only price is involved or coalitions of buyers where bundles of items are involved. In most of these approaches, payoffs can be split and shared, and they do not consider qualitative attributes like cities, types of accommodation, and so forth. Therefore, complex multi-attribute negotiations are not supported by current group buying approaches. Additionally, it should be noted that even though every buyer coalition may be considered a type of negotiation team at some point, not every negotiation team is a buyer coalition. For instance, let us imagine the negotiation between a union and the manager of an enterprise. The union may send a negotiation team formed by different experts or different stakeholders (i.e., representatives for different types of workers) to negotiate with managers. In this case, the goal is not obtaining volume discounts as group buyers' case. Moreover, buyer coalitions may be highly dynamic formations that may change when better coalitional options arise. That is not the case of the group of traveling friends or the union, where once the team has been formed; it usually remains static during the negotiation process.

\subsubsection{Cooperative Team Formation}

Another field relevant to this task is cooperative team formation \cite{cohen99,dignum00,gaston05,badica12}. When teams are formed, agents with different skills, necessary for achieving the common and shared goal, are sought. Team members with different skills/expertise may be desirable for complex negotiation domains. Nevertheless, cooperative team literature in multi-agent systems usually assumes that team members share the same goal and all of them work towards the achievement of such common goal. For that purpose, team members share an overall plan that they follow together, share knowledge of the environment, share intentions, and share skills needed to achieve the common goal \cite{sycara06}. 

Cohen et al. \cite{cohen99} present a framework where teams are groups of agents with a shared mental state that are committed to accomplish a joint goal. Speech acts are used as a mechanism to form and disband teams. If an agent requests another agent to perform a certain action, and it is followed by an accept message, both agents are considered to follow a joint goal and they cooperate for the realization of such goal. 

Dignum et al. \cite{dignum00} present a team formation framework based on persuasive dialogues and speech acts. An initiator agent forms an abstract plan. In the first stage, the agent gathers information on agents that may contribute to the plan attending to their abilities. Then, the initiator agent engages in persuasive dialogues with a subset of agents willing to work together towards the joint goal. The objective of the persuasion dialogue is to ensure that other agents are committed to the joint goal. Forming teams according to agents' skills should be a useful approach in situations where negotiation teams need different agent profiles to tackle the negotiation domain (e.g., experts in different areas, different services, etc.). Persuasion is also useful, especially in open environments where agents do not necessarily hold the same beliefs, goals, and intentions.

In \cite{gaston05}, a framework for dynamic team formation in a network of agents is presented. Each agent has a certain skill, and the agents are connected with a limited number of agents, forming a network. In this scenario, teams have to be formed dynamically in order to carry out tasks that are needed for the global goal of the network. Those agents that are not active can start the team formation process and look for teammates. However, an agent can only be part of a team if it is connected with at least one agent of the team. Networks are realistic structures in open systems, where agents do not know every single agent in the system, but a set of agents with whom it usually interacts. Thus, mechanisms such as the one presented in \cite{gaston05} offer realistic group formation principles for negotiation teams in open multi-agent systems.

Badica et al. \cite{badica12} present a framework for the formation of dynamic workflows of experts and services in environmental management applications. The organization of experts into workflows is carried out by  multi-issue negotiations that take into account the fact that experts may have previous commitments that limit their availability for team tasks. Negotiations are cooperative in the sense that the goal of all of the agents is a successful management in an environmental incident.

In most of the aforementioned frameworks, team members are usually fully cooperative with each other. Once agents commit to a joint goal, they cooperate and coordinate towards the consecution of that goal. Team members in a negotiation team may have different sub-goals or preferences despite sharing a common goal. Hence, team members may not be fully cooperative with other fellow team members, since other individual sub-goals may be at stake during the negotiation (e.g., trip conditions in the traveling friends' scenario). 

It should be highlighted that team formation algorithms are usually (not necessarily always) initiated by a single agent that asks for help to other agents on a specific goal/task. Hence, these mechanisms may be more appropriate in applications where a centralized authority does not exist and agents are more willing to fully cooperate.

\subsubsection*{\textbf{Special Challenges}}
The ideal formula for negotiation team formation may depend on the type of application that needs to be constructed. In any case, we argue that the aforementioned approaches may need to consider the following issues when forming a negotiation team:

\begin{itemize}

 \item Electronic commerce has given more social power to consumers, which now can find new sellers at a relatively low cost \cite{rezabakhsh06} and access information more easily. Not only that, but reputation models \cite{sabater05}  and gossiping \cite{younger05,perreau08} (i.e., exchanging information among known agents) may give an additional coercive power to consumers over sellers. In both cases, an opinion is formed about sellers which is based on the experiences/knowledge of the different agents in the society. This may produce sellers that are more willing to act cooperatively. Otherwise, sellers may be negatively rated by buyers. Thus, it is expected that, the larger the negotiation team, the greater social power it will be able to exert, and the more cooperative the seller will be.
 \item Even though it seems intuitive that the larger a negotiation team is, the better, this may not be always true. If the preferences of the team members are compatible and very similar, adding new team members to a negotiation team may only result in greater social power. As stated, if the preferences of the team members are very dissimilar, adding new team members may result in lower team performance \cite{sanchez-anguix13}. Thus, despite the fact that larger teams may be able to bring together more social power, intra-team conflict may deteriorate the quality of the final agreement to a point that greater social power does not compensate. Generally, it is not possible to exactly know the preferences of potential team members prior to the negotiation itself. Nevertheless, past experiences with negotiation partners \cite{brzostowski08} and recommender techniques like collaborative filtering \cite{schafer07} may help to accomplish the task of assessing which negotiation partners are more similar. In the former case, the agents use the history of past negotiations (negotiation outcome, and one's own utility function) with another agent to build a possibility distribution over the space of negotiation outcomes. Then, the expected utility of the negotiation with the agent is calculated and the agent negotiates with those agents whose expected utilities are higher. The same mechanism can be employed by one agent to determine which prospective teammates are the most similar. In the latter case, collaborative filtering consists of recommendation techniques for very large data sets. The rationale behind this technique is that if two agents have the same opinion on an issue, they are more likely to have a more similar opinion in other issues than randomly chosen agents. Thus, based on a history of past purchases/deals, it may be possible to determine whether or not two agents are similar. However, this information is usually provided/used by an authorized third party (e.g., an electronic marketplace).
 \item Related to the previous issue, more team members or stakeholders may bring additional negotiation attributes to the negotiation table. The first effect  over the negotiation is that the negotiation domain becomes larger and possibly computationally harder to work with. Despite this computational disadvantage, it may introduce negotiation attributes that are only interesting to a sub-group of the agents that participate in the negotiation. If the opponent is not interested in these attributes, it may make trade-offs easier. In contrast, if a negotiation attribute is introduced to satisfy a sub-group of the team members and the issue is important for the opponent, it may result in higher preference conflict with the opponent (i.e., a new issue which is important for the opponent and whose team preferences on issue values are opposite to those of the opponent). Hence, it may be more difficult to find an agreement. In general, additional issues in the negotiation may be a double-edged sword that can report both benefits and disadvantages.
 \item One of the problems that negotiation teams may face is tackling negotiation domains that are inherently complex. This means that the nature of the domain is hard to understand and it requires the expertise of different agents. For instance, when an organization negotiates in a complex negotiation, it  sends a negotiation team composed of different experts. These experts may come from the different departments of the organization (e.g., marketing, human resources, research \& development, etc.) and have different backgrounds that enrich the understanding of the problem. Information regarding agent roles and/or agent identities \cite{such11} may come handy to determine which potential team members are more fit for the negotiation problem. 
\item Different agents may provide different social relationships to the team's social network. Social networks may directly impact upon the performance of teams \cite{koc09} since it can provide with extra information for the team. They may provide new information and team members that are beneficial for the team.
\end{itemize}

\subsection{Opponent Selection}

Once the team has been formed, it is necessary to find suitable opponents from the list of prospective opponents. The team should decide which opponents they are going to face. If enough computational resources are available, all of the opponents can be selected and negotiations can be carried out in parallel. However, if computational resources are scarce, a subset of the  opponents has to be selected. 

\subsubsection{Opponent Evaluation Mechanisms}
As introduced earlier, each individual agent can calculate the expected utility of a negotiation with an opponent based on past outcomes and situations \cite{brzostowski08}. Another classic method to evaluate opponents is trust and reputation \cite{sabater05}. Trust is built as an evaluation mechanism based on one's own past transactions/interactions, while reputation is built based on the aggregated opinions of agents in the society. The rationale behind trust and reputation is that opponents that respect trade conditions (e.g., dispatch dates, payments, etc.) are positively evaluated, whereas agents that violate trade conditions (e.g., delays) or misbehave are evaluated negatively. Generally, agents should select opponents that are evaluated positively by trust and reputation mechanisms.

\subsubsection{Discussion in Groups}
The expected utility of negotiating with an opponent and trust evaluations are calculated from the point of view of one agent. Since the team is composed by more than a single agent, the opinions of different agents may need to be aggregated. In fact, there may be intra-team conflict since teammates' opinions on opponents may vary due to different experiences and preferences. In case of conflict, agents need to resort to other mechanisms for aggregating the opinions of teammates. One of these classic mechanisms is voting \cite{nurmi10}. Classic voting schemes like majority voting and Borda count have the advantage of aggregating opinions with a few interactions. 

However, they may not be expressive enough to determine the best option for the negotiation team. For instance, if one of the teammates has bad experiences with certain vendor but other teammates have good and more recent experiences, the team may determine to discard the opinion of the former team member. Deliberation and persuasion in groups can discover and clear these differences \cite{mcburney07,heras12}. Deliberation is a discussion to decide a course of action among a group of agents. McBurney et al. \cite{mcburney07} introduce the first formalization for deliberation dialogues in multi-agent systems. The interaction protocol for agents usually starts when agents have to decide on a course of action. Then, agents inform about the goals of the dialogue and the criteria that are relevant to the goals of the deliberation. After that, agents in the deliberation can enter a cycle where they are allowed to propose actions to be taken by the group, comment on actions proposed in the dialogue from different perspectives, revise the deliberation goals, recommend on actions that should be taken by the group, and confirm the acceptance of recommended actions. Heras et al. \cite{heras12} propose a persuasion framework for groups of agents that have to make a decision. In the framework, agents argue using case-based reasoning with two purposes. The first purpose consists of learning new solutions to domain problems, which are later used in dialogues to propose course of actions. The second purpose is learning how to argue from previous dialogues. 

\subsubsection*{\textbf{Special Challenges}}
Despite the fact that mechanisms exist that allow agents to evaluate opponents and discuss as a group on a course of action, as far as we are concerned it has not been applied to the evaluation of negotiation opponents. Comparing voting schemes and persuasion/deliberation, computation time may have a key on deciding on which technology should be chosen. If time is limited, voting schemes may be more appropriate when comparing a large number of negotiation opponents since they may require fewer messages. On the other part, if time is not of extreme importance in the process, deliberation and argumentation are more expressive and allow discovering scenarios like the one commented above.

\subsection{Understand the Negotiation Domain}

Understanding together the negotiation domain is a task of extreme importance. Not only does it allow team members to get a grasp of other team members' preferences, but it makes it possible for team members to tackle correctly the negotiation when the domain is complex and requires expertise in different knowledge areas. 

Even assuming that team members have similar backgrounds, it is still important to understand the negotiation domain together. Let us imagine that a group of friends (e.g., Alice, Bob and Charlie) decides to go on a travel together and have fun. What is the meaning of ``having fun''? Clearly, it may be different meaning for each friend: Alice thinks that if the city offers lots of adventure sports then the travel is fun, Bob thinks that having fun also involves finding a place with considerable night life, whereas Charlie is happy with any plan as long as it does not involve much money. From this situation, it can be inferred that the price, adventure activities included in the travel package, and the night life activities are relevant negotiation issues for the team in the negotiation at hand.

\subsubsection{Discussion in groups}

The technologies that can give support to these processes are varied. As stated, persuasion and deliberation in groups can help to manage conflict when deciding on a course of action \cite{mcburney07,heras12}. The object of discussion in this phase is those issues that are relevant for the team during the negotiation. Differently to the deliberation regarding opponent selection, in this case the available options and possible negotiation issues are not fully known beforehand. Thus, they have to be discovered during the argumentation process.

\subsubsection{Belief Merging}
 Other technologies like belief merging \cite{booth06,qi07,everaere07} may also prove useful for obtaining a shared model of the negotiation domain. Belief merging consists of joining together the knowledge that comes from different sources. If no inconsistency is present among the different data sources, information can be easily aggregated. However, if some piece of information is inconsistent among different data sources, special mechanisms have to be employed. 

Booth et al. \cite{booth06} present a belief merging algorithm for information coming from multiple data sources. It is assumed that one of the data sources is completely reliable (e.g., trivial information) with respect to the others. Consistent information is added as new knowledge, while inconsistent information undergoes a contraction process where it is incrementally reduced to pieces of consistent information by means of negotiation. For the application of this solution in a negotiation team, the sources of information can be considered the different agents in the team. The use of negotiation results convenient since in a negotiation team, each data source corresponds to a different party. The result of belief merging is information which can be used as shared mental model for the team. 

Qi et al. \cite{qi07} present another framework that merges different sources of information by means of negotiation among inconsistent sources. The main difference between this framework and the previous one is that the authors consider the fact that knowledge bases may have different priorities. These priorities represent different degrees of confidence in the different sources of information. This mechanism can represent the fact that in a negotiation team, the information of some team members may be more reliable in some areas.

 However, it should be noted that most belief merging methods are not strategy-proof \cite{everaere07,konieczny11}. A belief merging method is strategy-proof when it is robust against attempts of manipulation by agents. An agent may try to manipulate the belief merging process if it expects to increase its utility by acting that way. In \cite{everaere07}, the authors study the restricted conditions under which several families of belief merging operators are strategy-proof.  

\subsubsection*{\textbf{Special Challenges}}

Identifying negotiation issues that are relevant for the team is not the only task necessary to completely understand the negotiation domain. Identifying which attributes are predictable and compatible for the team, and which attributes are not predictable is also crucial. On the one hand, a negotiation issue is predictable and compatible among team members if the preferences of all of the team members over issue values are known and compatible. For instance, in a team of buyers, it is logical that all team members prefer low prices over high prices. In this type of negotiation issues, there is full potential for cooperation among team members since increasing the utility for a team member (i.e., decreasing the price) results in other team members staying at the same utility level or increasing their utility. On the other hand, a negotiation issue is not predictable among team members if nothing can be inferred about which issue values are preferred by team members. The issue may be compatible among team members (i.e., same ranking of preferences over issue values) or not, but it is not possible to know the nature of the negotiation issue unless team members are willing to share information. Using information regarding which issues are predictable and unpredictable among team members may be useful for deciding on which negotiation strategy is used among team members. For instance, in \cite{sanchez-anguix12} the authors present an agent-based negotiation team model that, as long as the negotiation domain is exclusively composed by compatible and predictable issues, it is capable of guaranteeing unanimity regarding team decisions at each round of the negotiation.

Another interesting issue is how team members analyze if they have incentive to share all of the information regarding the negotiation domain. An agent may be willing to share a piece information only if it expects that it is going to report higher utility than hiding the piece of information. Such et al. \cite{such12} propose disclosure mechanisms based on human psychology that allow agents to determine when sharing private information results in greater utility. However, research in this topic is still in early stages and it has not been applied to negotiation teams.

\subsection{Agree Negotiation Issues}

Since the previous stage produced a list of issues which are relevant for the team members, the next stage consists of agreeing a final list of negotiation issues with the opponent. The opponent may have its own list of issues relevant to the negotiation. Thus, a final list of issues to be negotiated should be agreed between both parties. 

From the initial set of negotiation issues proposed by the team, some of the issues may not be negotiable since the opponent does not offer that service. For example, if the team members had originally concluded that negotiating packages of adventure activities is a relevant issue to the team but a travel agency does not work with such packages, the issue cannot be included in the negotiation. Additionally, some negotiation issues that were not included in the list proposed by the team may be included in the final list since they are relevant to the opponent.

As for those negotiation issues present in the lists proposed by team members and the opponent, it may also be necessary to agree on the issue domain (i.e., the values that the negotiation issue can take). Similarly to the agreement on the list of issues, the final domain value may not contain all of the values proposed by both parties (i.e., Rome cannot be a value for the city of destination if the travel agency does not offer flights to Rome).

\subsubsection*{\textbf{Special Challenges}}

Despite being an important process in the pre-negotiation, very little attention has been paid to agreeing negotiation issues between parties. In fact, most researchers in negotiation assume that the list of negotiation issues and their domains are already agreed in their negotiation models. Faratin \cite{faratin00} mentions in his thesis the possibility of adding and removing non-core issues during the negotiation. While core negotiation issues remain static during the negotiation process, involved parties may be able to add or remove non-core negotiation issues as the negotiation process advances. However, the list of non-core issues is assumed to be known by both parties and the development of an issue-manipulation algorithm was appointed as future work. We acknowledge that this is a process that needs to be researched in the future.

 \subsection{Plan Negotiation Protocol}
 
 After the list of issues for the negotiation process is set, the different parties have to agree a negotiation protocol unless the protocol is already set by a third party or by the system. In that case, this task and the next task can be skipped. 

There are different negotiation protocols that may be applied for a specific situation. For instance, if the negotiation team engages with an opponent in a bilateral negotiation, both parties could employ the classical alternating offers protocol \cite{rubinstein82}, extensions of such protocol like the k-alternating offers protocol \cite{lai08}, protocols based on fuzzy tradeoffs \cite{luo03}, two-stage protocols based on optimization and concession \cite{pan12},  or complex mediated protocols like \cite{klein03b,ito08}.

 The team as a whole may have different opinions and knowledge about the available protocols. In fact, some of the team members may not even know some of such protocols. In that case, those protocols cannot be used by the team since some of its players do not know the rules and decision making strategies to face such games. 

\subsubsection{Discussion in Groups}
In this phase, if more than an applicable protocol is known by all of the team members, they should decide as a group which protocols are preferred by the team (i.e., a ranking of the known protocols). Again, the team may resort to voting \cite{nurmi10}, and argumentation and deliberation \cite{mcburney07,heras12}.
\subsubsection*{\textbf{Special Challenges}}
The main challenge in this task is how agents evaluate the different protocols available in the literature. As far as we are concerned, there is not an exhaustive study that analyzes which protocols are more convenient in different situations. Such a study would be necessary to provide agents with the mechanisms to make informed decisions on the negotiation protocol.

 \subsection{Agree Negotiation Protocol}
 
Considering that the team has already decided on which negotiation protocols are preferred by team members (e.g., some ranking over the negotiation protocols), they should agree with the opponents on the negotiation protocols that are to be followed for interacting. Again, opponents may not know how to play some games, making some of the options not feasible. Some protocols known by the opponent may not be known by all of the team members. Over the list of protocols that are known by both parties, both parties may have different preferences and knowledge regarding the different protocols. This decision between both parties is going to involve some kind of simple negotiation (i.e., we do not expect the number of possibilities to be large) or discussion among both parties. In some cases, besides the negotiation protocol, some parameters of the protocol have to be decided also by both parties (i.e., who is the initiating party in the alternating offers protocol \cite{rubinstein82}, the number of offers allowed in the k-alternating offers protocol \cite{lai08}, who acts as trusted mediator in mediated protocols like \cite{klein03b,ito08}, etc.). Despite not being explicitly covered by any work, we consider that most negotiation models can be easily adapted to this task.

\subsection{Decide Intra-team Strategy}

Negotiation protocols define the rules of interaction to be followed by the different parties. In a single player party, the decisions are individually taken by one agent. However, when the party is formed by multiple individuals, it is necessary to decide on \textit{how}, \textit{when}, and \textit{what} decisions are taken, and \textit{who} takes those decisions. This is what is termed as an intra-team strategy or negotiation team dynamics.

For example, in the case of the alternating offers protocol \cite{rubinstein82}, each party should decide on which offer is sent to the opponent party, whether or not to accept the offer proposed by the other party, and when one should withdraw from the negotiation process. Thus, any intra-team strategy for teams participating in the alternating offers protocol should decide on those issues. 

Obviously, for different negotiation protocols, different decisions have to be taken and an intra-team strategy that has been proposed for a particular negotiation protocol may not be directly applied to other type of negotiation protocol. Thus, usually negotiation protocols and intra-team strategies are tightly coupled.

\subsubsection{Simulation Studies}

One of our hypotheses is that there is not a single intra-team strategy that is capable of outperforming the rest of intra-team strategies for every possible scenario. Depending on the goal of the team (e.g., social choice performance measure), and depending on the conditions of the negotiation some intra-team strategies may perform better than others. By conditions of the negotiation we refer to those factors, either external  (e.g., the number of competitors, the number of prospective negotiation partners, etc.) or internal  (e.g., team size, similarity among team members' preferences, deadline length, etc.) that can affect the negotiation.  Usually, the number of variables is so large that it makes a theoretical analysis not feasible. Thus, some researchers have employed extensive simulations to assess which strategies would work better in certain conditions \cite{oliver96,matos98,tu00,gerding03}. Oliver et al. \cite{oliver96}, Matos et al. \cite{matos98}, Tu et al. \cite{tu00}, and Gerding et al. \cite{gerding03} propose evolutionary strategies in bilateral negotiations that converge towards reasonably good strategies in different environments. Nevertheless, some specific conditions of the negotiation have to be known beforehand to evolve appropriate strategies.

\subsubsection{Role Allocation}
 It may be the case that some intra-team strategies additionally require role/task assignment (e.g., information retrieval, monitoring the market, etc.). Nair et al. \cite{nair03} propose a framework that is capable of evaluating different role allocation and reallocation policies in cooperative teams by means of distributed Partially Observable Markov Decision Processes. This approach may be valid for intra-team strategies that require different roles. However, rewards steaming from role allocation may be uncertain in a negotiation team since agents are heterogeneous and they may not know each other. Additionally, it is a cooperative approach that does not take into account the preferences of agents over tasks. 

Hoogendorn et al. \cite{hoogendoorn06} present a negotiation framework for the allocation of tasks between agents. An agent can place requests for task distribution among other agents. Then, agents bid for being assigned to the tasks of their interest. A negotiated approach is more convenient when agents are heterogeneous and they have different preferences with respect to roles. However, in this approach only bids are considered to assign tasks to agents. It does not take into account other factors like past experiences, trust and reputation, etc.

\subsubsection*{\textbf{Special Challenges}}
Despite the fact that extensive simulations have been carried out in negotiations with single individual parties, no extensive study exists for the team case. The only exception is the work presented in \cite{sanchez-anguix12c,sanchez-anguix13}, where intra-team strategies are studied under varied negotiation conditions (i.e., deadline length, team member similarity, etc.). The work focuses on intra-team strategies where only predictable and compatible issues among team members are considered. Thus, it is acknowledged that more work is required in these lines.

With respect to role allocation, it should be pointed out that team members may also have their own individual goals. Teams with mixed motives have not been extensively studied in multi-agent literature, with the exceptions of \cite{grosz02,paruchuri07}. Grosz et al. \cite{grosz02} present a framework where agents have to reconcile conflict between team commitments and individual actions. If an individual action reports benefits and it is inconsistent with a committed team action, the agent can choose to decommit and pursue its own self-interest. Authors propose the use of social norms, with associated punishments and rewards, to make team commitments prevail. However, social norms are difficult to deploy in a negotiation team unless a central authority exists. Even so, it may not be possible to determine whether or not team members are collaborating since other agents' preferences are not known. Paruchuri et al. \cite{paruchuri07} present an extension of Partially Observable Markov Decision Processes to the team case. However, instead of modeling a single reward for the team, individual rewards are also modeled. This way, they can take into account agents' selfishness. The frameworks attempts to calculate optimal group policies that maximize the rewards of team members as much as possible. Nevertheless, rewards steaming from role allocation are difficult to anticipate. Therefore, it is necessary to further study to what extent team members would fully collaborate in negotiation team’s tasks in spite of their own utility. For example, how interesting is for a team member to look for new outside options for the team when current ones report high utility for himself? A self-interested team member may decide to neglect its search tasks and continue with present outside options if it considers that new outside options will not increase its current welfare. We consider that this is a promising line of work that needs more attention.

\subsection{Select Individual Strategy}

Each team member should plan its individual strategy before heading into the negotiation. An intra-team strategy defines mechanisms for team decision-making but they do not define how individual team members behave when playing those mechanisms. It is up to the agent to decide how to act inside the team: it can be more or less cooperative. The agent should also decide its attitude with the opponent. The two aforementioned factors will define the initial negotiation strategy of each team member.

Generally, the selection of the initial negotiation strategy is based on what is expected about the opponent and teammates.  As stated in the previous section, one of our hypotheses is that the conditions of the negotiation environment play a key role in selecting which intra-team strategies are more appropriate for each specific situation. Thus, team members should also decide on their individual strategy based on the knowledge about conditions of the negotiation. We consider that the mechanisms employed in other domains involving single player parties are also applicable in this task. Hence, no special challenge arises.

\subsection{Negotiation \& Adaptation}

The final phase is the negotiation itself. During this phase, team members should follow the planned intra-team strategies, individual strategies, and negotiation protocols. However, negotiation is a dynamic process that may not work as planned (e.g., opponents not behaving as one initially thought, team members performing below/above one's expectations, team members leaving the team, etc.). Therefore, it may be necessary that each team member adapts its own negotiation strategy, and that the team replans some of the aspects related to team composition and team dynamics. 

\subsubsection*{\textbf{Special Challenges}}
We argue that some of the special challenges that arise in a negotiation team during the negotiation are:

\begin{itemize}
 \item Team membership: As stated, team membership may be dynamic. In fact, how dynamic a negotiation team is may depend on the domain application. Domains where team members are more self-interested, and less bonds exist between team members (e.g., team of buyers) may result in more dynamic negotiation teams than domains where team members are more cooperative and there are human bonds behind team membership (e.g., group of travelers, human organizations, etc.). In any case, in both situations the problem of dynamic membership may arise. For instance, new buyers may appear in the electronic market that may join the team to take advantage of larger price discounts. Similarly, a new traveler decides to travel during holidays and his user states the desire of joining the pre-existing group of travelers. Cases of team member's withdrawal are also possible. For instance, one of the buyers participates in other buyers' coalitions and decides to close a deal, making its membership in the rest of the buyers' coalitions no longer necessary. 
 
\item Negotiation issues: As stated, both parties agreed to negotiate over some initial issues. For some reasons (e.g., computational issues, computational tractability, etc.), they may have decided to leave some less relevant issues out of the negotiation table. However, at some points of the negotiation an impasse \cite{spector95,babcock97} may occur in the negotiation. A negotiation impasse occurs when the parties are unable to reach an agreement and the perspectives of reaching one are very negative. They are in a deadlock. A possible solution for such problematic situation is what is known as issue linkage \cite{tollison79,sebenius83,morgan90}. Basically, when parties negotiate on one issue, adding another issue and linking its value to the value of the initial issue can increase the probability of finding an agreement. The new issue may be added to reduce intra-team conflict  (e.g., how costs are split in the team), or they may be added to reduce conflict with the opponent (e.g., include a payment method issue and maximize the preferences of the opponent in the new issue). This adaptation heuristic during the negotiation may be positive for cases that are prone to fail. However, as suggested by \cite{sebenius83,morgan90}, issue linkage may also have negative effects since it may also reduce the agreement space. As of today, issue linkage is an area that has not been widely studied in automated negotiation, where it has been assumed that issues remain static during the negotiation process. Hence, it is an area that requires further exploration, especially for the team case since conflict may appear at the team level and the opponent level. 

\item Intra-team strategy and individual strategy adaptation: As stated, intra-team strategies define \textit{what} decisions are taken by the team, \textit{how} decisions are taken, and \textit{when} those decisions are taken. In case that some negotiation conditions change, it may be wise for team members to change their current intra-team strategy to match accordingly the new changes. Obviously, changes in the intra-team strategy and environment's condition also call for an adaptation in team members' individual strategies. In this sense, there have been some works that advocate for a change in individual agents' strategies in automated negotiation. We can distinguish between works where individual agents adapt their behavior attending to environmental conditions like outside options and competitors \cite{sim03,li06,an08,ren09,an10,ren11} and works where individual agents adapt their behavior during the negotiation attending to the attitude of the opponent \cite{robu05,robu08,narayanan06,hindriks08,williams11}. 

Sim et al. \cite{sim03} introduce market driven agents.  These agents adapt their concession strategies based on market conditions like eagerness, trading time, trading opportunities and trading competition. Later, Ren et al. \cite{ren09} improve market driven agents by considering not only current market options, but also future opportunities. Differently from other approaches, in \cite{ren11} the authors propose bilateral negotiation strategies for multi-issue negotiations that account for market conditions. A stochastic Markov chain is proposed in An et al. \cite{an08} with the goal of modeling market dynamics in single-issue negotiations.  In their model, they assume that agents have a probability distribution over future trading opportunities, reservation prices, and deadlines. A sit-and-wait strategy is also assumed so at the first negotiation step both partners offer their highest utility value and no new bid is offered until the first private deadline, where the agent offers its reservation value. Opponent offers are accepted based on the prediction of new and better offers in the future. Li et al. \cite{li06} present a bilateral negotiation framework where the reservation price is updated based on the information of current trading opportunities and future trading opportunities. In \cite{an10}, the authors propose a bilateral negotiation model with decommitment for leasing resources on the cloud. The negotiation strategy of buyers takes in consideration environmental factors like the demand ratio of resources, the cost of resources, and market competition for resources. 

Robu et al. \cite{robu05,robu08} present a learning mechanism used by sellers to adapt to the preferences of buyers. The preferences are represented by utility graphs, which model interdependencies between binary issues. Narayanan et al.\cite{narayanan06} present a negotiation framework where pairs of agents negotiate over a single issue (i.e., price). The authors assume that the environment is non-stationary in the sense that agents' strategies may change over time. Non-stationary Markov chains and Bayesian learning are employed to tackle the uncertainty in this domain, and eventually converge towards the optimal negotiation strategy. Hindriks et al. \cite{hindriks08} present a negotiation framework for bilateral multi-issue negotiations where agents' preferences are represented by means of additive utility functions. The main goal of this work is learning a model of the opponent's preferences (i.e., issue value rankings, and issue weights), and Bayesian learning is used for this purpose. Williams et al. \cite{williams11} present a strategy that is capable of adapting its concession strategy based on the predicted behavior of the opponent. 

The previous works adapt different elements of the negotiation (e.g., preference models, strategies, reservation prices, etc.). However, they usually focus on modeling aspects of the environment where the negotiation is being carried out or the opponent. Both opponent's models and environment's models can be applied to the team case. However, there is an additional layer of negotiation/interaction, which in this case is carried out within teammates. One agent should also consider the actions and behaviors of its teammates in order to act accordingly in the negotiation. Thus, new learning and adaptation mechanisms based on the information that is generated inside the team by team members are needed. 
\end{itemize}

\section{Discussion \& Conclusions}
\label{conclusions}

Agent-based negotiation teams is a fairly novel topic in the area of multi-agent systems and artificial intelligence. More specifically, an agent-based negotiation team is a group of interdependent agents that join together as a single negotiation party because of their shared interests at the negotiation process. The reasons to employ these multi-individual parties are: (i) more computation and parallelization capabilities; (ii) unite agents with different expertise and skills whose joint work makes possible to tackle complex negotiation domains; (iii) the necessity to represent different stakeholders or different preferences in the same party (e.g., organizations, countries, married couple, etc.). In this article, we have analyzed and reviewed the tasks that may help agent-based negotiation teams to perform successfully. 

Being a special type of automated negotiation where at least one of the parties is a group of agents, it inherits some of the challenges from the classic topic of automated negotiation like opponent selection, typical pre-negotiation tasks (e.g., setting negotiation issues, agreeing negotiation protocol, etc.), strategy selection, and so forth. Nevertheless, an additional layer of difficulty is added to those challenges due to the fact that team members may have different expertise and different preferences (i.e., intra-team conflict). Hence, those decisions that once were taken alone or with the opponent parties should now be discussed and agreed within the negotiation team. Moreover, some unique challenges like negotiation team formation and deciding the intra-team strategy (i.e., team dynamics) that should be employed in the negotiation also arise with the new topic.

In this article we have organized, in a workflow, the tasks that a negotiation team should take. The workflow aims to be general and adaptable to a wide variety of domains and applications. For each of these tasks we have highlighted the aforementioned challenges that arise with the field of agent-based negotiation teams and we have pointed out to related fields of study that may hold the right answer for such challenges.   Next, we briefly describe the  workflow and the challenges that we have identified throughout this article. 
\begin{itemize}
 \item \textbf{Identify Problem}: The first step consists of being able to perceive when conflict is present, and who may be involved in the conflict (i.e., opponents, prospective teammates, and competitors). For that matter, it is necessary to analyze the environment and look for potential partners. The literature has largely focused on domain dependant conflict detection and search mechanisms in networks and markets, leaving conflict detection in open systems almost unexplored. This latter aspect is relatively important for making general negotiators in a wide variety of domains. Hence, this area of work remains largely unexplored. 

 \item \textbf{Team Formation}: If the agent thinks that it may be beneficial to form a negotiation team, it should attempt to select its teammates. Closely related research areas are coalition formation, buyer coalitions, and cooperative team formation.  On the one hand, coalitional approaches (including buyer coalitions) have focused on forming optimal groups of agents and how to divide the payoff generated by the group task. However, the result of the negotiation with the opponent may be difficult to anticipate making it complicated to estimate the value of coalitions. Moreover, while the payoff generated by some negotiation issues like price may be naturally divisible, the payoff generated by other issues is not divisible (e.g., payment method). On the other hand, cooperative team formation aims to form teams for a certain task based on complementary skills. The usual trend in the literature is to consider these team members as fully cooperative and committed to the team goal. Nevertheless, in agent-based negotiation teams, team members may not be fully cooperative since they may additionally have their own and possibly conflicting individual interests. Special challenges identified in this task are considering (i) the relationship between team size and social power; (ii) the relationship between team size, team similarity, intra-team conflict and conflict with the opponent; (iii) requiring different knowledge expertise; (iv) the social network provided by each team member.

 \item \textbf{Opponent Selection}: The next step consists of selecting the opponents with whom the team will negotiate. Classic approaches to this task have relied on trust and reputation, or the negotiation outcome in previous interactions with the opponent. Since the aforementioned perspectives are usually individual, it is necessary to form an opinion for the team. In this case, we pointed out that social choice and argumentation/deliberation in groups can play this role. The main challenge in this workflow task is adapting the aforementioned mechanisms to the opponent evaluation task.

 \item \textbf{Understand the Negotiation Domain}: The general idea behind this task is creating a shared model of the negotiation domain at hand. It includes identifying relevant negotiation issues, merging different points of views and expertise, clarifying team goals that may be abstract in essence, and identifying the nature of prospective negotiation issues (e.g., compatible and predictable, unpredictable, etc). Discussion in groups can certainly help to discover these issues during argumentation processes. Additionally, belief merging is also a potential candidate to solve this task as long as team members are truthful or the belief merging method is strategy-proof. The main challenges in this task are identifying the nature of negotiation issues among team members, and taking into account opportunistic behaviors (e.g., manipulating belief merging, hiding relevant information for one's own interest, etc.).

\item \textbf{Agree Negotiation Issues}: The next part consists of agreeing with the opponent which negotiation issues should be considered in the negotiation. Depending on the participating parties, it is possible that some of the issues/issue values are not available for negotiation (e.g., the service provider cannot offer such issue). Additionally, it is also possible that some issues/issue values that were not initially considered by the team are included due to opponent parties' requests.  Despite its importance, negotiation models usually assume that the negotiation domain is given and they do not provide mechanisms that allow forming or negotiating the domain of each negotiation. This is an interesting research challenge since, as far as we know, there is almost no related literature.

\item \textbf{Plan Negotiation Protocol}: Given a specific situation, there may be different negotiation protocols that may be used to negotiate with the opponent. Team members should argue about which protocols are preferred according to their experiences, known strategies , and so forth. Although the problem has not been explicitly covered by the literature, its solution may not pose exceptional efforts compared with classic argumentation and social choice problems. Nevertheless, it is also acknowledged that there is a lack of works comparing the advantages and disadvantages of available negotiation protocols under different conditions.

\item \textbf{Agree Negotiation Protocol}: Similarly, once team members have discussed about the available negotiation protocols, they should negotiate a proper negotiation protocol and its parameters with the opponent. Current negotiation technologies may be able to support this phase.

\item \textbf{Decide Intra-team Strategy}: Intra-team strategies define team dynamics for a specific negotiation protocol. They define the coordination and negotiation protocol carried out within the team to decide on the steps to be carried out in the negotiation with the opponent. Team members may employ information regarding the current environment state (e.g., deadline length, number of competitors, team size, beliefs regarding the opponent, etc.) in order to decide on the most appropriate intra-team strategy. For this matter, it may be necessary to carry out simulation studies in order to discover under what circumstances some intra-team strategies perform better than others. The scarce research in negotiation teams has gone in that direction. We have identified that even though some studies exist that identify good practices and good strategies for single individual parties, the area remains largely unexplored for intra-team strategies. Moreover, if the intra-team strategy requires role differentiation, techniques from role/task allocation may be employed. However, it should be considered that agents may not be fully cooperative, which is the usual approach in role/task allocation. Thus, the problem slightly differs from classic role/task allocation. 

\item \textbf{Select Individual Strategy}: Intra-team strategies define team dynamics, but they do not define the individual behavior of team members per se. The next step consists of each team member individually deciding the most appropriate individual behavior for the negotiation at hand. This task may not pose additional difficulties compared with the selection of the individual negotiation strategy in classic negotiation literature.

\item \textbf{Negotiation \& Adaptation}: The final task consists of negotiating and adapting in order to face unexpected events properly. We have identified three special challenges. The first of them is team membership adaptation. The second challenge that needs some consideration is adapting negotiation issues. Parties can solve impasses in the negotiation and better off other parties by including other issues that were not initially included in the negotiation. As far as we are concerned, this problem has not been widely studied in multi-agent literature. The third and final challenge that we consider is the adaptation of the intra-team strategy and the individual strategy. Usually, team members have planned on using an intra-team strategy and an individual strategy based on some initial prediction of the negotiation environment, opponents, and teammates' behavior. However, based on new evidence, initial predictions may prove wrong and adjustments need to be done in order to properly tackle the negotiation. In automated negotiation, some works exist that allow single individual parties to adapt themselves to changes in the negotiation environment and new predictions. However, these mechanisms need to be adapted for the team case since interactions among team members also influence the negotiation.

\end{itemize}

 Therefore, it is observable that some issues remain open and they need to be solved before deploying full-fledged applications that make use of agent-based negotiation teams. In this article, we have observed that despite these open issues, technology exists that can be adapted in the near future to solve the aforementioned challenges. We expect that, due to inherent social component of negotiating as a group, the deployment of applications supporting agent-based negotiation teams can specially benefit from social networks. For instance, travel services are classically offered to individuals in Web sites. Even though it is possible to purchase travel services for multiple individuals, usually the process only involves interactions with one single user, leaving the intra-team negotiation, if any, to human users. Human negotiations may be time consuming and costly in cognitive terms. A social network for travelers that introduces computational models for negotiation teams may help to avoid these problems. In this social network, users may indicate to their personal agents their desire to go on a travel together. Then, the agents help users from the initial stages of forming a group of travelers (i.e., a negotiation team) to the later stages where the group of agents looks for appealing travel agreements with different travel agencies. The general scheme of this social application can be extrapolated to other domains involving purchases in group.

 Another future application that may be supported by agent-based negotiation teams is management of complex democractic organizations. For example, agricultural cooperatives are supposed to be democratic institutions where groups of farmers join together to save resources for the distribution of their products. One of the main problems of agricultural cooperatives is the principal-agent problem \cite{ortmann07}. Basically, despite being democratic institutions, agricultural cooperatives are managed by a board of directors who take decisions on behalf of the democratic institution. It has been reported in the literature \cite{ortmann07} that dissatisfaction in cooperatives comes from the fact that the goals of members are not aligned with those of the managers. Agent-based negotiation teams may provide support for part of the processes that are carried out by cooperatives. For instance, the negotiations between agricultural cooperatives and distributors may be supported by an electronic market where the agricultural cooperative is modeled as an agent-based negotiation team. Each member may be represented by an electronic and personal agent that participates in the negotiation team according to the preferences of its owner. For this kind of application, it is specially important to provide scalable models for agent-based negotiation teams.

 We hope that with this article, the reader can be introduced to the novel topic of agent-based negotiation teams and that it facilitates their work towards building successful agent-based negotiation teams. As of today, our work has focused on designing new intra-team strategies for negotiation teams. We plan to extend our present research with new intra-team strategies, and additional phases of the proposed workflow like team formation based on agents' preferences.

\section*{Acknowledgements}
This work is supported by TIN2011-27652-C03-01 and TIN2012-36586-C03-01 of the Spanish government.

\bibliographystyle{elsarticle-num} 
\bibliography{references}

\end{document}